\documentclass[a4paper,fleqn,usenatbib]{mnras}
 \pdfoutput=1

\usepackage[T1]{fontenc}
\usepackage{ae,aecompl}

\usepackage{graphicx}	
\usepackage{pdflscape}  
\usepackage{amsmath}	
\usepackage{amssymb}	

\title[ALMA observations of dusty early-type galaxies]{ALMA observations of massive molecular gas reservoirs in dusty early-type galaxies}

\author[A.E. Sansom et al.]{\parbox{\textwidth}{
A.E. Sansom$^{1}$\thanks{E-mail: AESansom@uclan.ac.uk},
D.H.W. Glass$^{1}$,
G.J. Bendo$^{2}$,
T.A. Davis$^{3}$,
K. Rowlands$^{4}$,
N. Bourne$^{5}$,
L. Dunne$^{3}$,
S. Eales$^{3}$,
S. Kaviraj$^{6}$,
C. Popescu$^{1}$,
M. Smith$^{3}$,
and S. Viaene$^{6,7}$
}
\vspace{0.4cm}
\\
\parbox{\textwidth}{$^{1}$Jeremiah Horrocks Institute, School of Physical Sciences and Computing, University of Central Lancashire, Preston, Lancashire, PR1 2HE, UK\\
$^{2}$Jodrell Bank Centre for Astrophysics, School of Physics and Astronomy, The University of Manchester, Oxford Road, Manchester M13 9PL, UK\\
$^{3}$School of Physics and Astronomy, Cardiff University, Queens Buildings, The Parade, Cardiff CF24 3AA, UK\\
$^{4}$Department of Physics and Astronomy, Johns Hopkins University, Bloomberg Center, 3400 N. Charles St., Baltimore, MD 21218, USA\\
$^{5}$Institute for Astronomy, University of Edinburgh, Royal Observatory, Edinburgh EH9 3HJ, UK \\
$^{6}$Centre for Astrophysics Research, University of Hertfordshire, Hatfield, Hertfordshire, UK\\
$^{7}$Department of Physics and Astronomy, Krijgslaan 281, S9, 9000 Gent, Belgium 
}}

\date{Accepted XXX. Received YYY; in original form ZZZ}

\pubyear{2018}

\begin{document}
\label{firstpage}
\pagerange{\pageref{firstpage}--\pageref{lastpage}}
\maketitle

\begin{abstract}
Unresolved gas and dust observations show a surprising diversity in the amount of interstellar matter in early-type galaxies. Using ALMA observations we resolve the ISM in z$\sim$0.05 early-type galaxies. 
From a large sample of early-type galaxies detected in the Herschel Astrophysical Terahertz Large Area Survey (H-ATLAS) we selected five of the dustiest cases, with dust masses M$_d\sim$several$\times10^7$M$_\odot$, with the aim of mapping their submillimetre continuum and $^{12}$CO(2-1) line emission distributions. These observations reveal molecular gas disks. There is a lack of associated, extended continuum emission in these ALMA observations, most likely because it is resolved out or surface brightness limited, if the dust distribution is as extended as the CO gas. However, two galaxies have central continuum ALMA detections. An additional, slightly offset, continuum source is revealed in one case, which may have contributed to confusion in the Herschel fluxes. Serendipitous continuum detections further away in the ALMA field are found in another case. Large and massive rotating molecular gas disks are mapped in three of our targets, reaching a few$\times10^{9}$M$_\odot$. One of these shows evidence of kinematic deviations from a pure rotating disc. The fields of our two remaining targets contain only smaller, weak CO sources, slightly offset from the optical galaxy centres. These may be companion galaxies seen in ALMA observations, or background objects. These heterogeneous findings in a small sample of dusty early-type galaxies reveal the need for more such high spatial resolution studies, to understand statistically how dust and gas are related in early-type galaxies.
\end{abstract}




\section{Introduction}



 The interstellar medium (ISM) in early-type galaxies (ETGs - including E and S0) can give clues to their past evolution. While the ISM is not always such a large proportion of the mass in ETGs as it is in spirals, the state of the ISM can be a relic of the history of ETGs, which helps to identify how they evolved (Xilouris et al. 2004). Misalignments of molecular gas and stellar morphologies and kinematic position angles indicate an external origin for gas in some ETGs (Davis et al, 2011). The presence of offset gas and dust, away from the galaxy centre, also indicates that these did not originate from within the galaxy itself (e.g. Hibbard \& Sansom 2003; Donovan, Hibbard \& van Gorkom 2007; Serra et al. 2012; Kaviraj et al. 2012; Russell et al. 2014). On the other hand, ETGs do not have a random distribution of kinematic misalignments and many local ETGs show good alignments, as discussed in Eales et al. (2017). This may be due to an initially random distribution becoming increasingly aligned as the gas and dust settles in the gravitational field of the galaxies, or it may point to an internal origin for the gas and dust in some ETGs. Timescales for gas and dust to relax into rotating disks, aligned with the optical photometric and kinematic axes in ETGs, are typically expected to be quite short in the central few kiloparsecs (a few dynamical times or within $\sim10^8$ years, Tohline et al. 1982; van de Voort et al. 2015). Exceptions can occur however, if accretion from external cold gas is long-lived, as shown in simulations by van de Voort et al. (2015), lasting up to $\sim$2 Gyrs, 

 Many studies have shown that a cool ISM is present in a large fraction of ETGs, in the form of neutral gas (e.g. Welch, Sage \& Young, 2010; Oostereloo et al., 2010), molecular gas (e.g. Knapp \& Rupen 1996; Young et al. 2011; Alatalo et al. 2013; Young et al. 2014) and dust (e.g. Goudfrouij \& de Jong 1995; Temi et al. 2004; Finkelman et al. 2012; Smith et al. 2012; Rowlands et al. 2012), but not in all ETGs. Evidence of cold gas in ETGs also comes from the presence of star formation seen in UV emission (Kaviraj et al. 2007, 2011). There is evidence that ISM phases in ETGs do not correlate well: dust-to-gas ratios and gas proportions per stellar mass show larger variations in ETGs than they do in spirals (e.g. Morganti et al. 2006; Combes, Young \& Bureau 2007; Welch, Sage \& Young 2010; Young et al. 2011; di Serego Allighieri et al. 2013; Lianou et al. 2016) and hot plasma phases are present in different amounts depending on aspects such as age (Sansom et al. 2006) or mass and location (Su et al. 2015; Anderson et al. 2015) of the ETG. Observed correlations between ionized gas and dust masses in ellipticals are weak and there is a lack of correlation with stellar masses (e.g. Kulkarni et al. 2014 and references therein). In spiral galaxies dust is ubiquitously associated with molecular gas but studies of these different ISM components are needed to understand how gas and dust are related in ETGs.

Recently, the Herschel Space Observatory observed many ETGs during its 3 year mission, particularly in the H-ATLAS survey (Eales et al. 2010). The H-ATLAS survey{\footnote {H-ATLAS survey, http://www.h-atlas.org/ }} is a submillimetre survey in five wavebands, using both the PACS and SPIRE detectors. It covers five areas on the sky and was obtained between 2009 and 2013 during the Herschel mission. A description of data release 1 for the three equatorial field regions is given in Valiante et al. (2016). H-ATLAS revealed larger amounts of dust in ETGs than previously expected (Agius et al. 2013), for galaxies selected purely on their optical morphologies. In this current work we look at five dusty examples from the H-ATLAS survey with new ALMA observations. The aim is to investigate where the dust and molecular gas are located and the relation between these components in these galaxies. This first paper presents the ALMA maps and basic results found. A future paper (Glass et al. in preparation) will look in more detail at the derived properties and their interpretation. 

In this paper, Section 2 describes the parent sample and sample selected for ALMA observations. Section 3 describes the ALMA observations and reductions. Results are presented in Section 4 and discussed in Section 5, whilst a summary is given in Section 6. We assume a Hubble constant of $H_0=70$ km s$^{-1}$ Mpc$^{-1}$.


\begin{figure*}
	\hspace{-1.40cm}\includegraphics[angle=0,width=190mm]{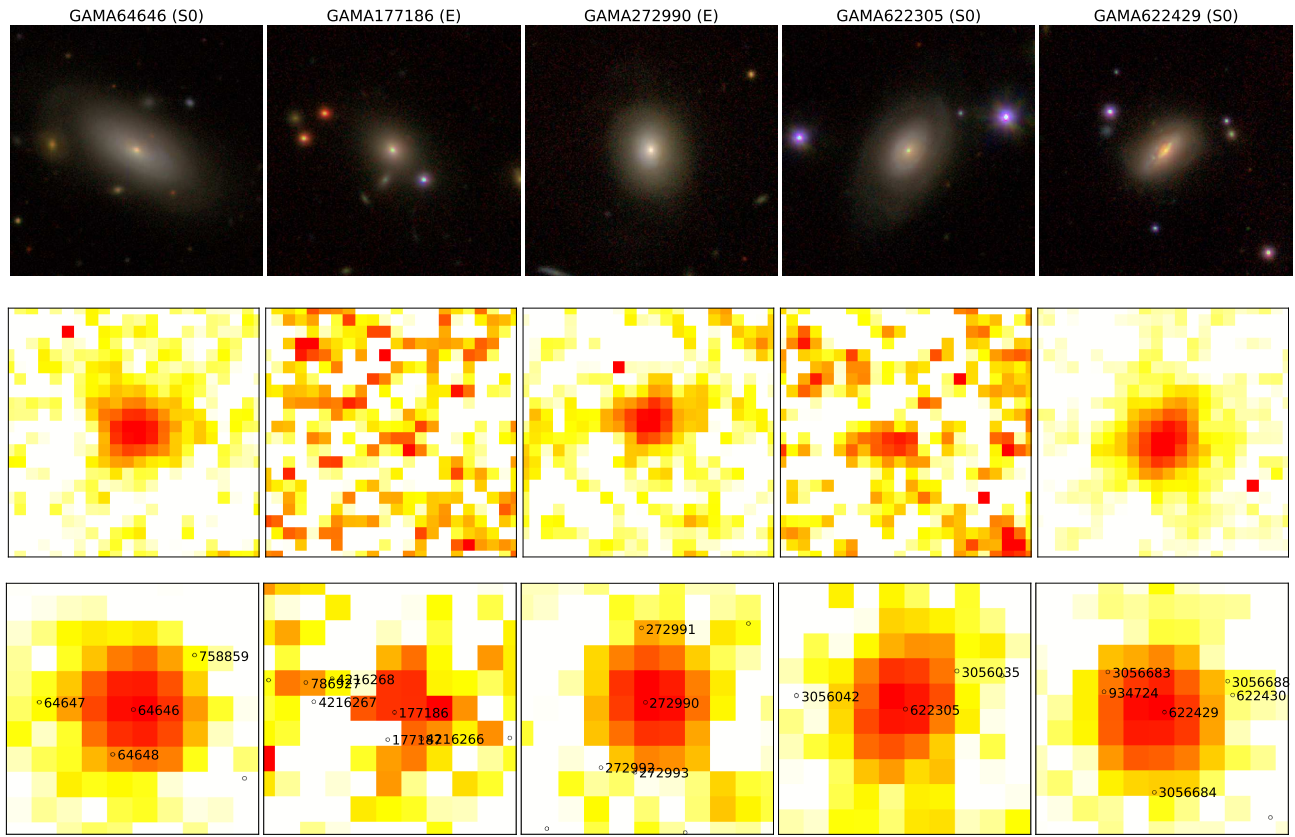}
\vspace{-2.6cm}
    \caption{All images are approximately 1 arcminute by 1 arcminute, with north at the top and east on the left. Upper row:  
KiDS BGR composite images of these five galaxies (processed by Lee Kelvin at ARI/LJMU to a common zero point and gain). Middle row: H-ATLAS PACS 100$\mu$m images (3\arcsec\ per pixel). Lower row: H-ATLAS SPIRE 250$\mu$m images of these five galaxies (6\arcsec\ per pixel), illustrating their strong submm detections in H-ATLAS, centred on the optical galaxy. GAMA IDs are indicated in these fields.}
    \label{fig:opt_hatlas_figure}
\end{figure*}

\begin{figure}
	\includegraphics[width=110mm]{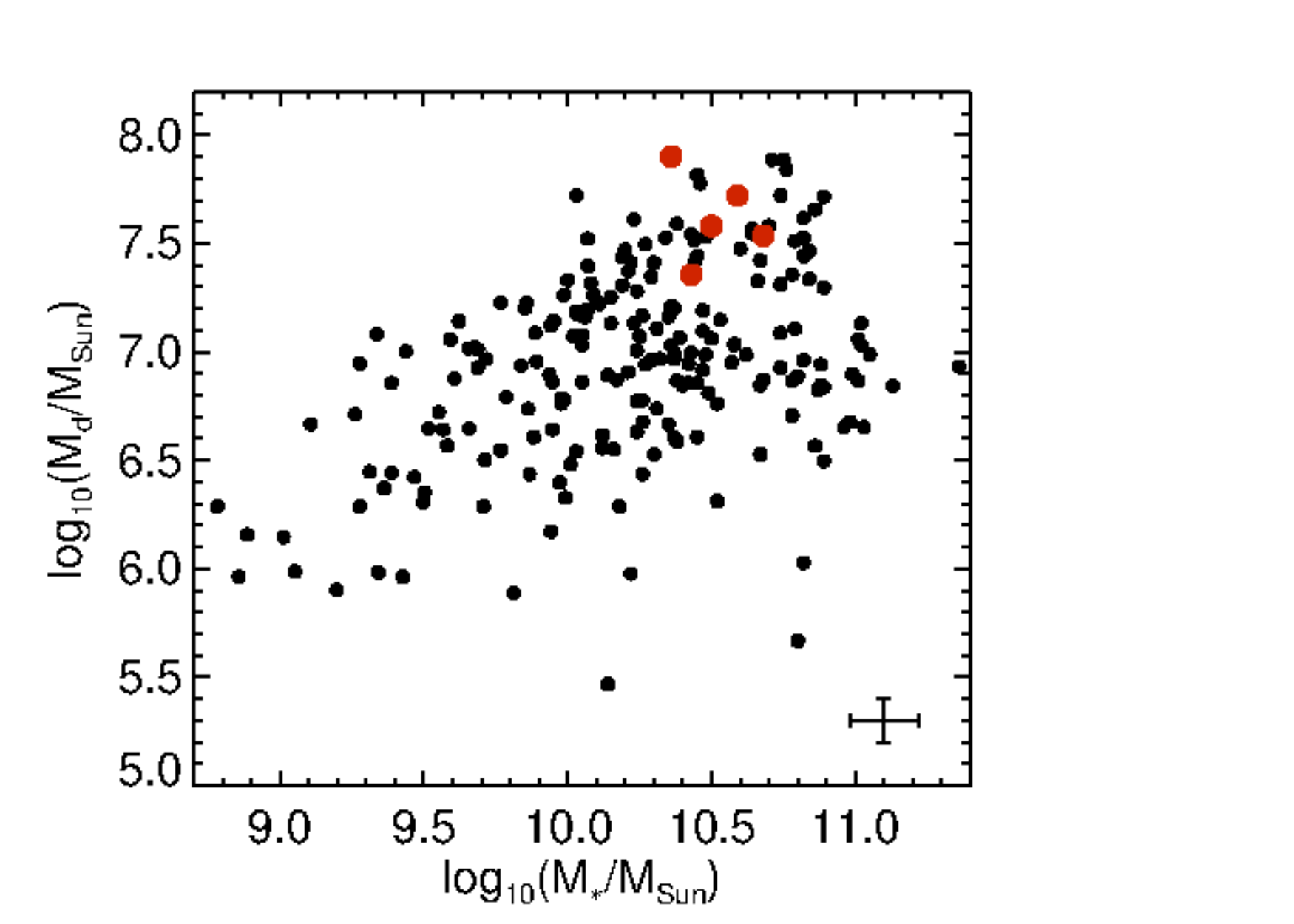}
    \caption{Log of dust mass, $Log(M_d)$, plotted against log of their stellar mass, $Log(M_*)$, for the parent sample of H-ATLAS detected ETGs (from Agius et al. 2013), in solar units. Our five target galaxies with ALMA observations are highlighted as larger filled red symbols.}
    \label{fig:Mstar_Mdust_figure}
\end{figure}

\section{Sample Selection}

 The H-ATLAS survey overlaps with the Galaxy and Mass Assembly (GAMA) survey (Driver et al. 2011), which provides multi-waveband characteristics and redshifts for all galaxies down to $r<19.8$ mag. Three GAMA/H-ATLAS regions near the equator cover a total of $\sim$160 deg$^2$ on the sky and were the first to be extensively studied and classified.
To obtain a census of dust content in ETGs in the local Universe we compiled a complete sample of ETGs in the three GAMA equatorial fields at $0.013<z<0.06$ and separated them into submm detected and undetected samples based on H-ATLAS observations. See Agius et al. (2013) for full details of this H-ATLAS parent sample of ETGs.
In summary we studied a complete optical sample of 771 nearby ETGs in the H-ATLAS equatorial fields and detected 220 of them at 250$\mu$m. These nearby GAMA galaxies are generally in low density environments (Agius et al. 2015).
Optical morphologies were based on Sloan Digital Sky Survey (SDSS) images classified in Kelvin et al. (2014), plus removal of some objects observed to have spiral structure. After removal of galaxies with flattening greater than E6, emission line active galactic nuclei (AGN) and gravitational lens candidates, this left 220 galaxies in the submm detected parent sample (SubS). Of these, 188 had detections above 3-sigma in at least 3 of the 5 H-ATLAS bands, giving good measures of their dust temperatures and masses (assuming a single modified black body spectrum in the submm).

 From this SubS parent sample we selected five of the apparently dustiest cases for mapping with ALMA observations. These five were selected to have dust mass estimates from H-ATLAS of $>2\times 10^7$M$_\odot$, and whose optical morphology showed ETG structure (E or S0/a from Kelvin et al. 2014). They were chosen to have r-band half-light radii of $\sim$ 4\arcsec$<R_e<10$\arcsec\, leading to predicted sizes of gas and dust distributions that were expected to match well with the spatial dynamical range available with ALMA. To make predicted size estimates for the dust and gas we had to assume a disc structure similar to that observed in some ETGs, with sizes typically $R_e/2$ or $\sim$ 1 or 2 kpc (e.g. Young et al. 2011; Davis et al. 2013). However, these predicted sizes are uncertain and a major motivation for ALMA observations is to find out if the dust and gas are in relaxed structures or more irregularly distributed in dusty ETGs. We selected targets with 0.035$<$z$<$0.048, to both limit their extents and to help ensure their optical morphological classifications. Heliocentric redshifts (from NED), along with r-band half-light radii and positions from the GAMA survey are listed in Table 1.

 Optical and H-ATLAS images of our 5 target galaxies are shown in Figure 1. These optical images show generally smooth light distributions typical of elliptical and lenticular galaxies. There are no strong dust lanes seen in absorption in these optical images. Three were classified as S0/a (GAMA64646, GAMA622305, GAMA622429) and two as E (GAMA177186, GAMA272990) in the GAMA survey, by Kelvin et al. (2014), based on SDSS images. These classifications are the same in the latest visual morphologies catalogue in GAMA data release 3{\footnote {GAMA DR3 is at http://www.gama-survey.org/dr3/ }}. The top row in Figure 1 shows 
deeper images from the recent Kilo-Degree Survey with the VST (KiDS, de Jong et al. 2017). There is a lack of spiral structure. The deeper and sharper optical mages from KiDS, which are available through the GAMA consortium collaboration, show quite smooth spatial distributions for these galaxies, but reveal disturbed morphology near the centre of GAMA622429, with evidence of a dust lane. Optical centres are uncertain by $\sim\pm 0.7$\arcsec\, (from NED{\footnote{NASA Extragalactic Database, https://ned.ipac.caltech.edu }} position error ellipse values). H-ATLAS absolute pointing positions are uncertain by $\sim\pm$ 1.0\arcsec\, to 2.4\arcsec\, depending on signal-to-noise ratio (Maddox et al. 2017). The beam size for H-ATLAS at 100$\mu$m is 11.4\arcsec\, at 250$\mu$m it is 18\arcsec\, FWHM (Valiante et al. 2016; Valtchanov 2017). While Figure 1 shows that there are other GAMA galaxies in the field, the submm emission from H-ATLAS is mainly associated with the optical target galaxies rather than other GAMA galaxies. This does not preclude non-GAMA objects as potential contamination sources in the submm (Karim et al. 2013). 
  
The H-ATLAS PACS data, with a beam size of 11.4 and 13.7\arcsec\, FWHM at 100 and 160$\mu$m respectively (Valiante et al. 2016), show very little extent for our targets. PACS 100$\mu$m data are shown in the middle row in Figure 1, with 3\arcsec\, pixels. Gaussian fits to those data give extents comparable to the expected PACS beam size, with GAMA64646 being the largest at 12.8\arcsec\, FWHM. They are slightly non-circular, as expected for PACS 100$\mu$m point sources observed in the H-ATLAS fast scanning mode (Valiante et al. 2016). SPIRE 250$\mu$m data are shown in the lowest row in Figure 1, with 6\arcsec\, pixels.

 H-ATLAS dust masses $(Log(M_d))$ are plotted, together with the SubS parent sample, in Figure 2, illustrating the high dust masses and range of M$_*$ values of the current five ALMA targets. Dust masses plotted in Figure 2 are from modified black body fits to the submm data in Agius et al. (2013) and stellar masses are from the GAMA survey (Taylor et al. 2011).

\begin{table*}
\setlength{\tabcolsep}{4pt}
	\centering
	\caption{Observed ETGs, redshifts, r band $R_e$ radii, ALMA Band 6 exposure times, pointing positions (from CASA listobs files), smallest angular size major axis (BMAJ), minor axis (BMIN), position angle (BPA) and largest angular sizes (from antennae baselines) are listed. Calibrators are listed in the right column in this table. }
	\label{tab:table1}
	\begin{tabular}{lcccccccccc} 
		\hline
ETG   &      z &  $R_e$    &  Exp  &    RA    &          Dec &           & SAS       &       & LAS       & Calibrators \\
GAMA  &        &           &       &          &              &  BMAJ     & BMIN      &  BPA  &           & \\   
ID    &        & (\arcsec) & (sec) &  (J2000) &    (J2000)   & (\arcsec) & (\arcsec) & (deg) & (\arcsec) & Phase,       Flux and Bandpass \\
		\hline
64646 &  0.036898 & 9.7 & 1216.2 & 14h38m16.50s & -00d20\arcmin59.1\arcsec &   0.71 & 0.60 & 65.4 & 10 & J1448+0402 J1337-1257 J1550+0527 \\ 
177186 & 0.047628 & 3.8 &  501.6 & 11h44m11.75s & -01d51\arcmin56.4\arcsec &   0.72 & 0.64 & 51.3 & 11 & J1150-0023  J1229+0203 \\
272990 & 0.041133 & 4.2 & 1550.6 & 12h06m47.23s & +01d17\arcmin09.8\arcsec &   0.74 & 0.69 & 33.0 & 11 & J1220+0203  J1229+0203 \\
622305 & 0.042584 & 4.6 &  846.7 & 08h50m12.38s & +00d39\arcmin24.6\arcsec &   0.69 & 0.64 & -19.8 & 11 & J0839+0104  J0750+1231 \\
622429 & 0.040943 & 4.4 &  895.1 & 08h53m37.44s & +00d44\arcmin09.7\arcsec &   0.69 & 0.64 & -19.7 & 11 & J0839+0104  J0750+1231 \\
		\hline
	\end{tabular}
\end{table*}

\section{ALMA Observations and Reductions}

 Five dusty ETGs were observed with Band 6 receivers in ALMA Cycle 3 (Project code $=$ 2015.1.00477.S, PI$=$Sansom, configuration C36-3). Details of these observations are given in Table 1. Typically 40 of the 12m antennae were used, with baselines from $\sim$15m up to $\sim$640m. The receiver was tuned to include the $^{12}$CO(2-1) line at 230.5 GHz, in one of four spectral windows available for each target observation. Each spectral window is 1875 MHz, with two windows either side of the oscillator frequency. Three of the spectral windows have 128 channels covering 1875 MHz with a resolution of 31.25 MHz ($\sim$40 km/s). The fourth spectral window covering the CO(2-1) line has 3840 channels covering 1875 MHz, but the raw data are Hanning smoothed{\footnote {See ALMA Technical Handbook, Section 5.5.2}} to a resolution of 976.563 kHz ($\sim$1.27km/s). This set-up was used to allow observations of both continuum and line emission from each target. RMS noise levels were aimed to be around $\pm0.03$ mJy in the continuum and around $\pm3$ mJy in the CO line. Phase calibrators were observed every $\sim$7 minutes. Flux calibrators were also observed as well as bandpass calibrators for calibrating the waveband throughput and shape. These observations and calibrators are detailed in Table 1.

 Raw ALMA visibility data were calibrated using standard calibration scripts specifically created for each Execution Block. Reductions were done using CASA version 4.6.0, or 4.7.2 for GAMA64646 (https://casa.nrao.edu/docs/cookbook/). The reference antenna used was DV11, which is fairly central in the array. Checks were made for irregularities in the phases and amplitudes of the visibility data as a function of time and channel (frequency), using {\it plotms} software. Additional flagging of one antenna (DV02) was done for GAMA272990 and GAMA622305, and further flagging was done for the baseline calibrator for GAMA64646, which showed a sharp phase shift around 0h20m on 15 May 2016.

 Cleaning was first carried out using CASA software (using the same versions of CASA as above) for most of the field, with a spectral binning factor of 16 for the spectral window containing the expected CO(2-1) line emission, yielding a channel width of 10.16 km/s. The spatial binning was 0.12\arcsec\, square per pixel, which samples the beam well (see Table 1). 
We initially produced image cubes with no continuum subtraction so as to identify whether any CO line emission was present. When line emission was detected, we identified the channels with solely contiunuum emission and used these channels as input into CASA routine {\it uvcontsub} to remove the continuum from the visibility data. Final CO image cubes were made from these continuum-subtracted data by manually cleaning the image cube frame-by-frame, selecting spatial regions of line emission with a polygon mask. Continuum images were created using continuum channels from the spectral window with the CO line (or all channels from this window if no line was detected) and all of the channels from the other three spectral windows. Natural weighting was used when cleaning to optimize the detection of large-scale structure. This yielded synthesised beams of $\sim$0.65\arcsec\, FWHM (for a full list of beam sizes see Table 1). The typical equivalent beam area in pixels is $\sim 35$, which is the factor needed to convert from flux per pixel to flux per beam.  After imaging, primary beam corrections were applied. The continuum images cover 45.7\arcsec\, across and are cut off at 0.1 of the central primary beam efficiency.
Moment maps were generated in CASA and also using our own routines.

\section{Results}

  To make the CO maps shown below we used the makeplots IDL software{\footnote {https://github.com/TimothyADavis/makeplots}}, which utilizes the masked-moment technique (see e.g. Davis et al. 2018). In brief this technique uses a smoothed clean cube and defines a mask above a user specified threshold in the cube. Figure 3 shows CO line emission and velocity  maps for three of our five ETGs, showing the emission above a level of 1.5 times the rms noise. The lowest row in Figure 3 shows the spectra. Figure 4 shows CO line emission for one other source in each of the fields of GAMA177186 and GAMA622305, which are offset from the target galaxies. The lowest row in Figure 4 shows the spectra. The synthesised beam size is shown in each plot. Thus we detected some CO emission in all five fields. The continuum images are shown in Figure 5. The centre of each image is the optical centre of the galaxy, as listed in Table 1. These ALMA observations have a pointing uncertainty of approximately $\pm0.035$\arcsec\, (see https://help.almascience.org).
The noise levels in the data were approximately as expected.

\begin{figure*}
\includegraphics[width=96mm, angle=-90]{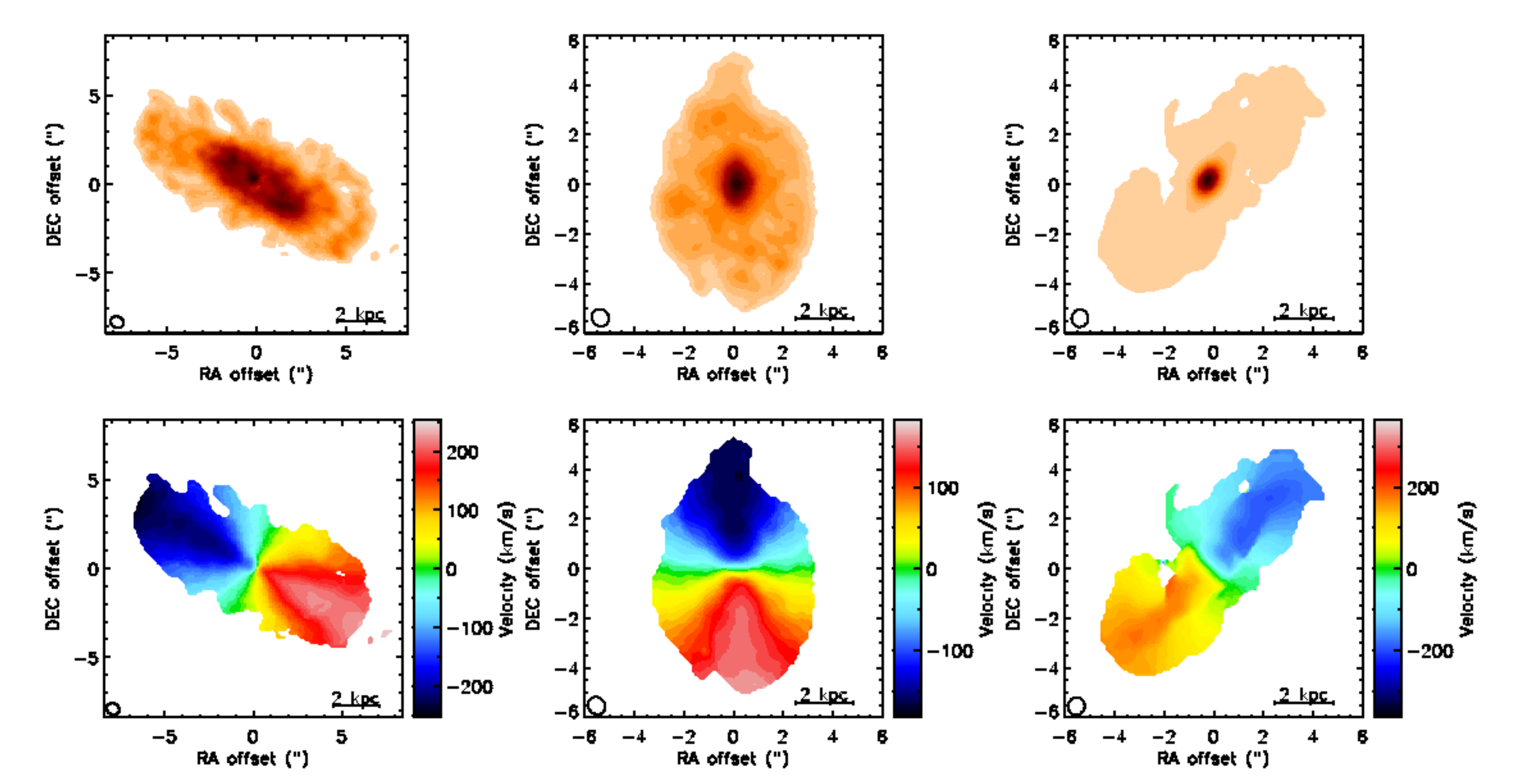}
\includegraphics[width=58mm]{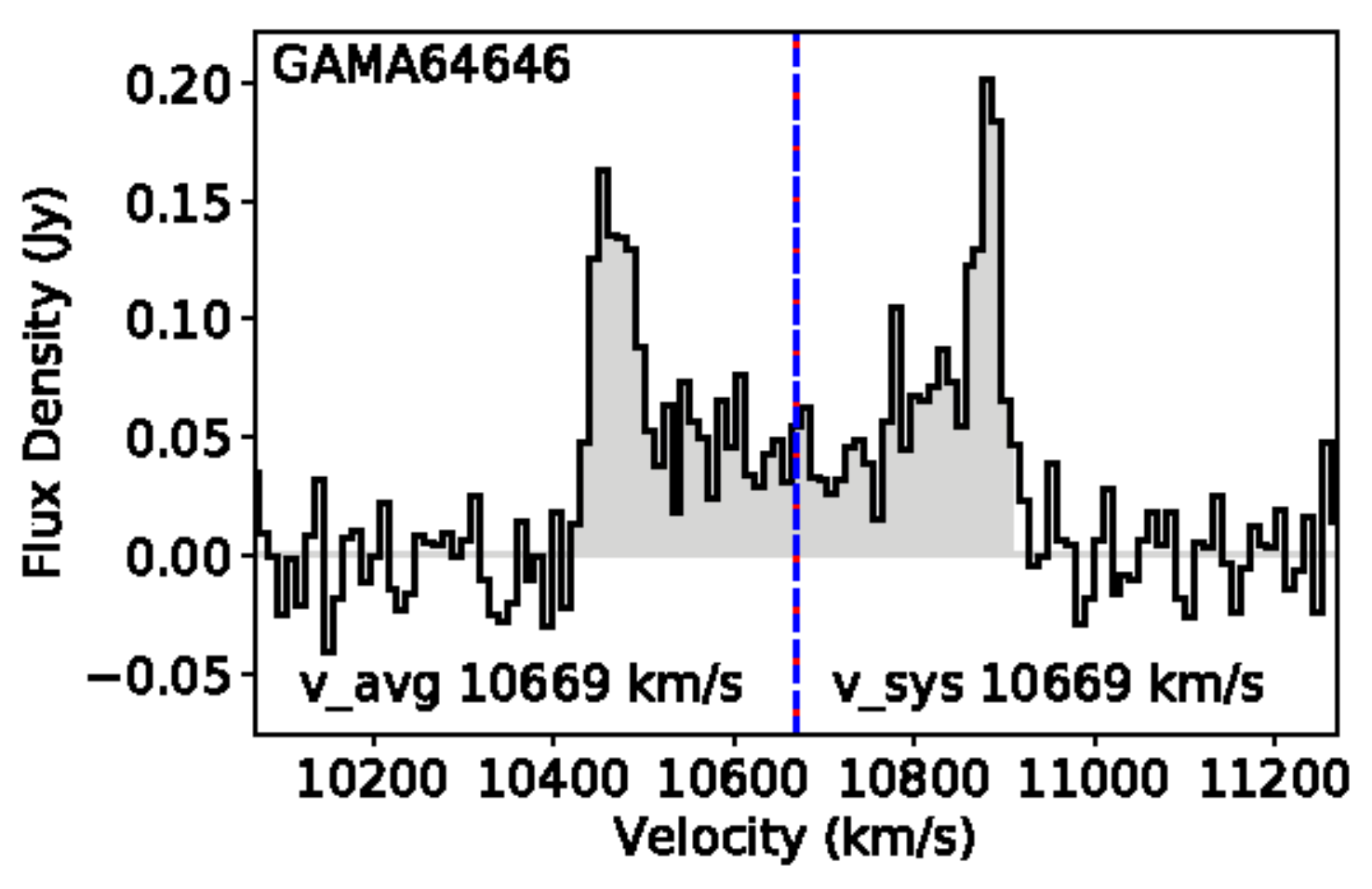}
\includegraphics[width=58mm]{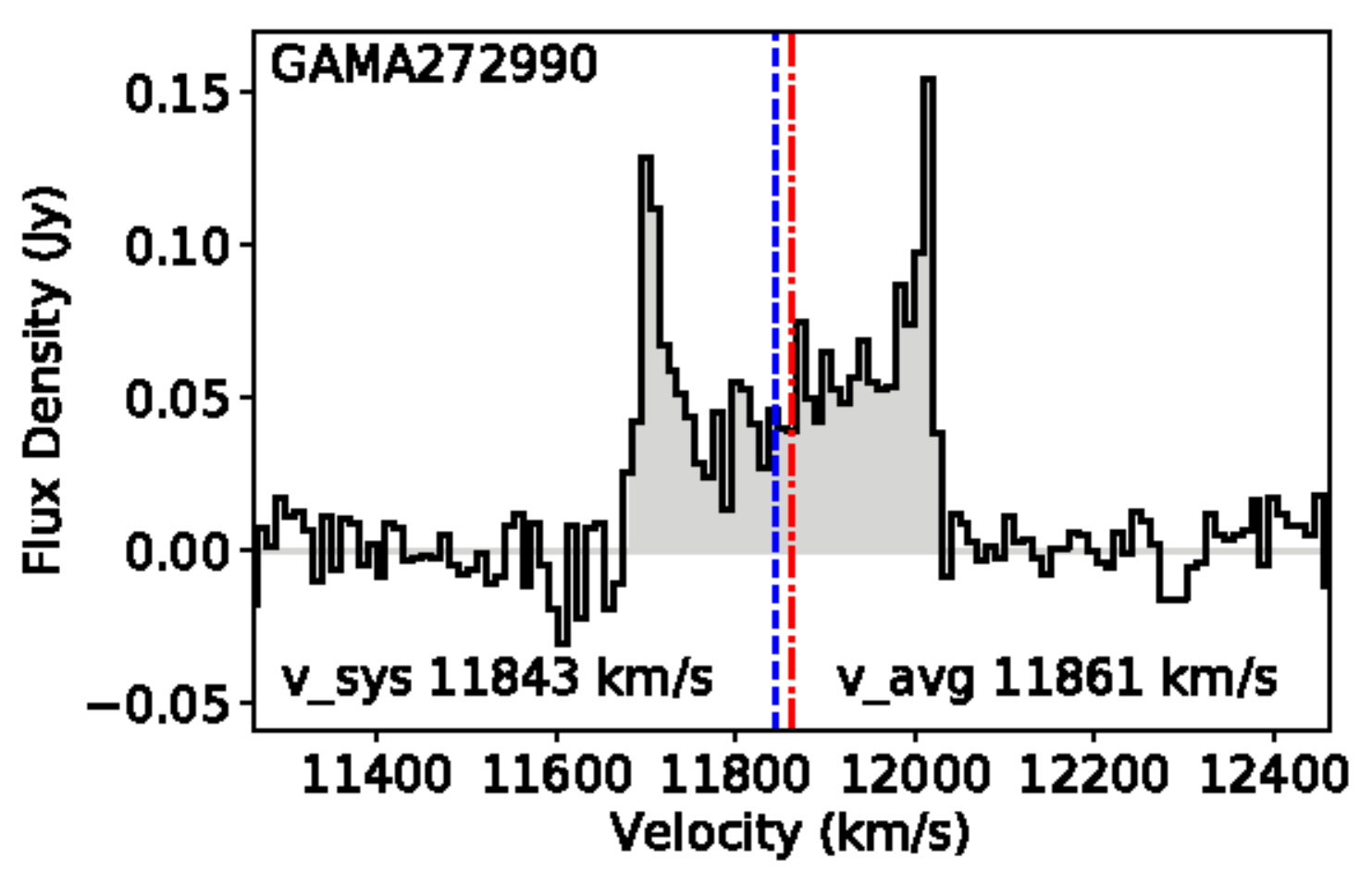}
\includegraphics[width=58mm]{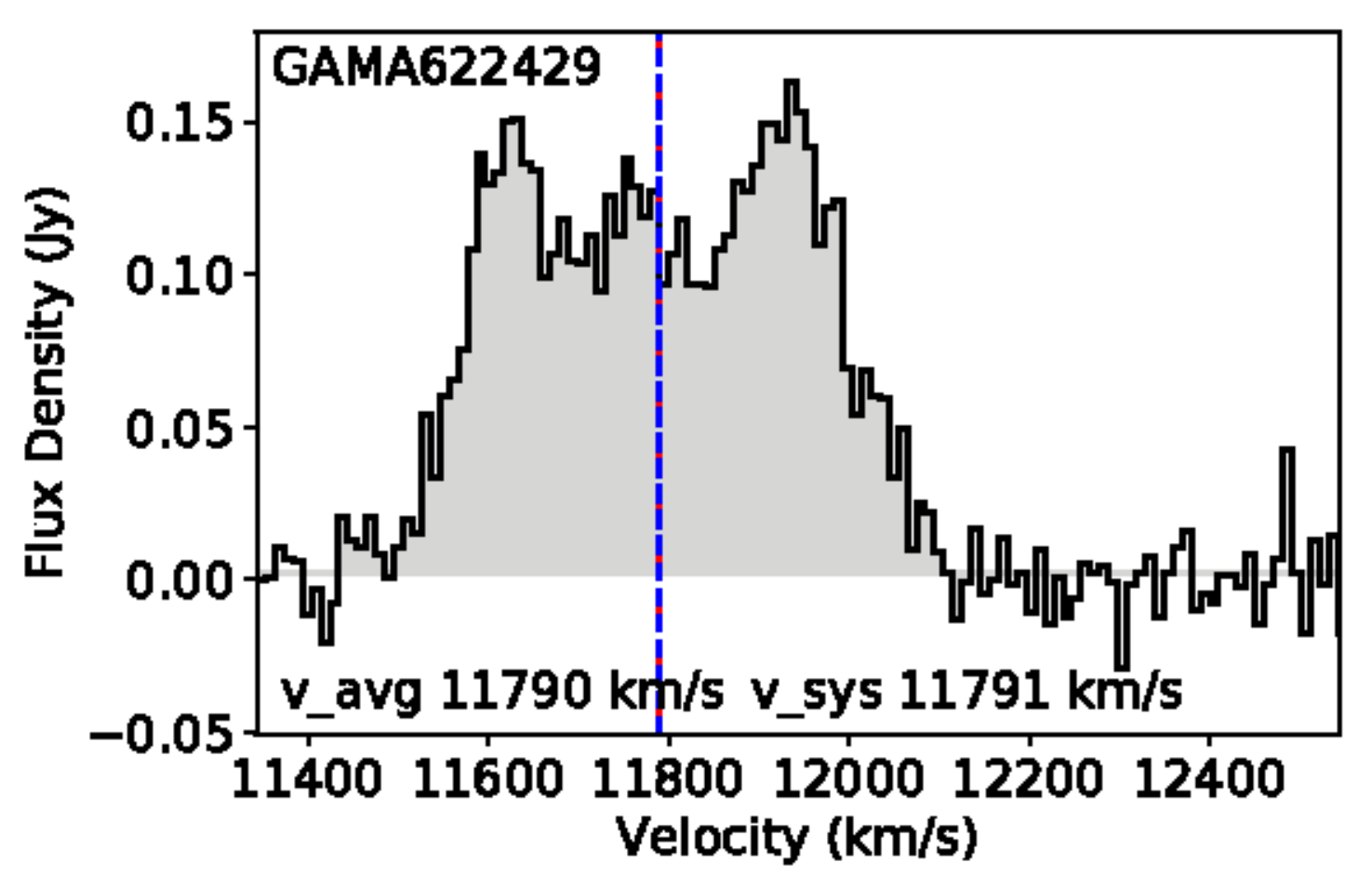}
   \caption{Moment maps and spectra for GAMA64646, GAMA272990 and GAMA622429 from left to right. {\it Upper row:} $^{12}$CO(2-1) line emission images from ALMA Band 6 central observations, with scale shown. Regions with fluxes above 1.5 $\times$ the rms noise are shown. {\it Middle row:} $^{12}$CO(2-1) first order moment maps showing velocities of gas in detected target galaxies, again with scale shown. {\it Lower row:} Spectra of the $^{12}$CO(2-1) emission line in each galaxy. The red dot-dashed line is the mean velocity for the channel range selected and the blue dashed line is the systemic velocity of the target galaxy in each case.}
    \label{fig:ALMA_figures}
\end{figure*}

\begin{figure*}
\includegraphics[width=124mm]{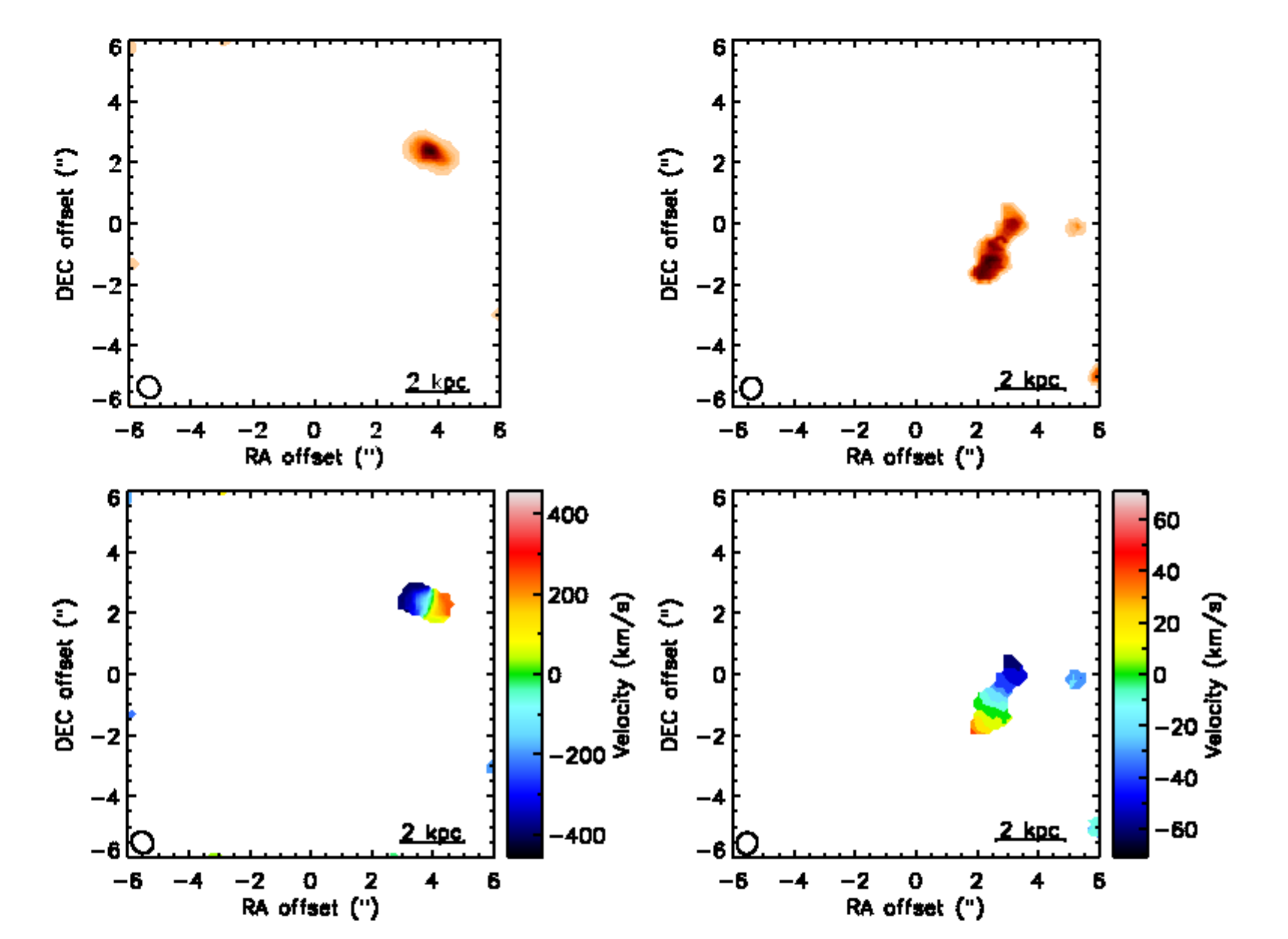}
\includegraphics[width=58mm]{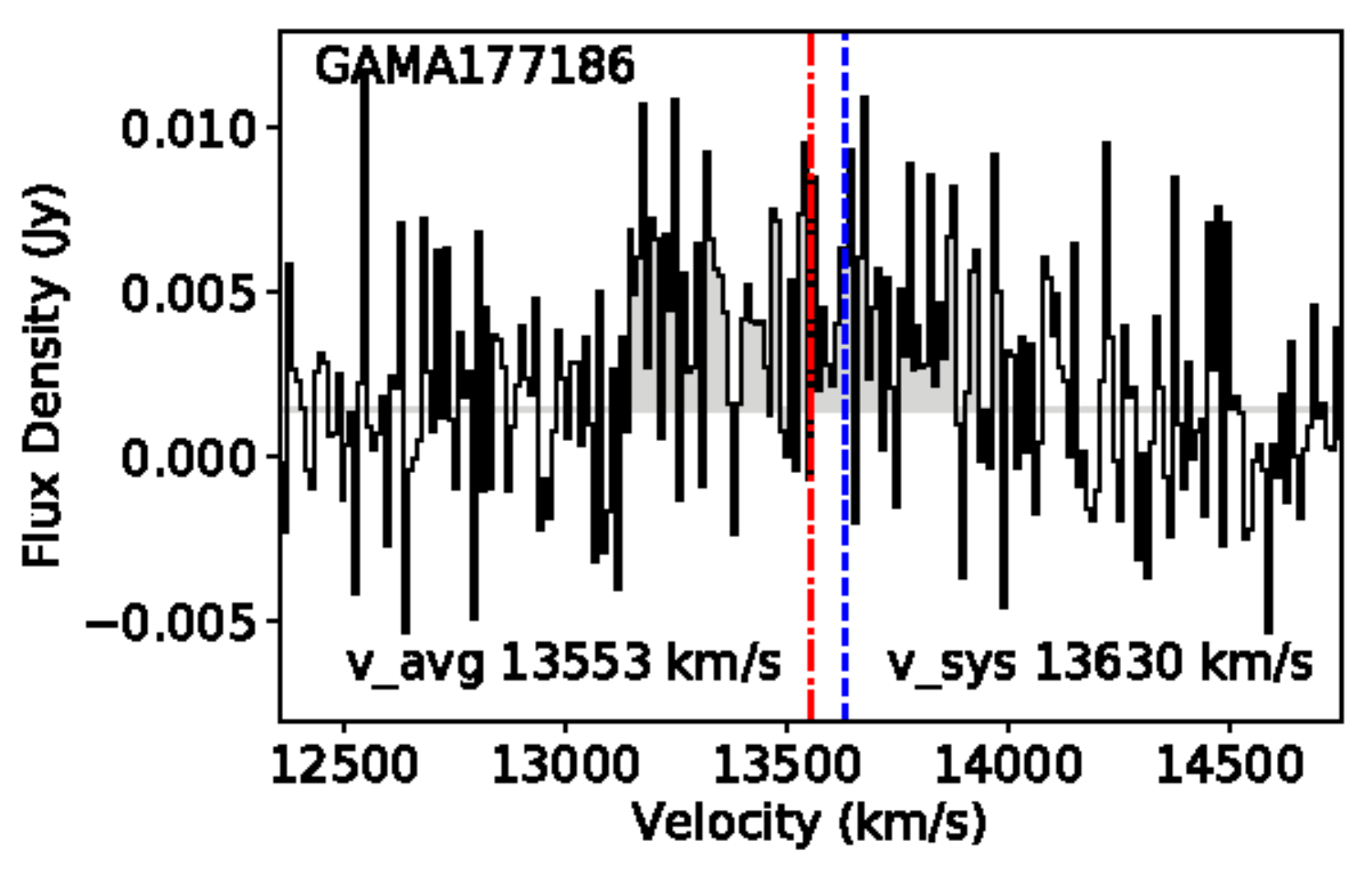}
\includegraphics[width=58mm]{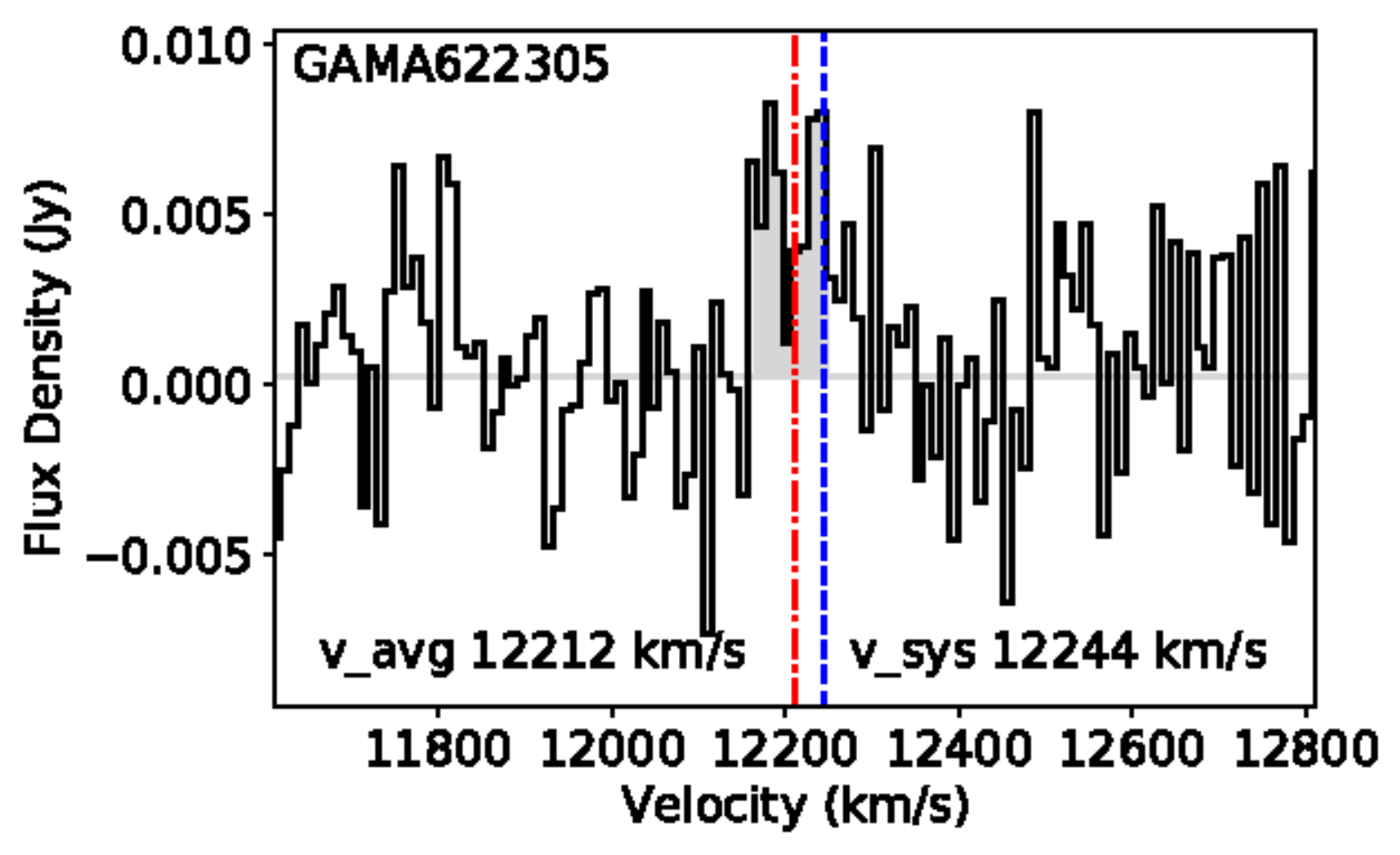}
   \caption{$^{12}$CO(2-1) line emission moment maps from ALMA Band 6 observation centred on GAMA177186 (left) and GAMA622305 (right), with scales shown. These images ({\it upper row}) reveal weak line emission detections offset from the target galaxies, but no line emission from GAMA177186 or GAMA622305 themselves. Regions with fluxes above 1.0 $\times$ the rms noise are shown. First order moment maps ({\it middle row}) show rotation in these offset sources. {\it Lower row:} Spectra of the $^{12}$CO(2-1) emission line region. The red dot-dashed line is the mean velocity for the channel range selected and the blue dashed line is the systemic velocity of the target galaxy in each case.} 
    \label{fig:ALMA_C0_G177186_G622305_offsets}
\end{figure*}

\begin{figure*}
\includegraphics[width=178mm]{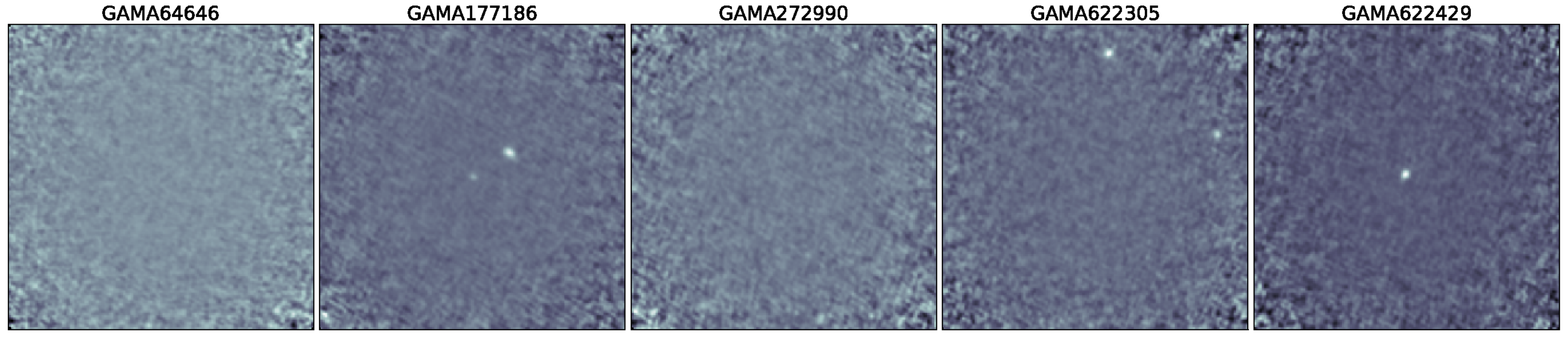}
   \caption{Continuum submm images from line free regions of the ALMA spectral windows. These continuum images from CASA have been corrected for the primary beam and show a subset of the ALMA field, covering the central 34\arcsec\, by 34\arcsec\,. Each field is centred on the target galaxy.}
    \label{fig:ALMA_Continuum}
\end{figure*}

\vspace{0.5cm}
Radio velocities (Optical velocities/(1+z)) are given below, unless otherwise stated. The images and cubes show the following features for each target:\\

GAMA 64646:
Continuum image shows no detection. CO data cube shows that emission is clearly detected and extends across more than 18\arcsec\, (13.8 kpc) in diameter. It is a rotating molecular gas disk as seen in Figure 3
(detected in velocity range 10420 to 10918 km/s).
See Appendix for a plot of the channel maps for this extended CO source (Figure A1). \\

GAMA 177186:
Continuum image shows a weak central detection, plus stronger detection nearby, at 4.4\arcsec\, north-west of the target galaxy. This offset source appears slightly elongated. CO data cube shows no detection apparent in the channel maps but a weak detection shows up in the offset source when the cube is collapsed, especially in velocity range $\sim$ 13121 to 13985 km/s. We confirmed the absence of line emission outside of this range by further binning of the channels.
Moment maps and the spectrum are shown in Figure 4 left, illustrating the source location and rapid rotation in the offset source. 
This offset source has a broad emission line whose centre is only 77 km/s below the systemic velocity of GAMA177186, if the line is from $^{12}$CO(2-1), which would put it at a similar distance to GAMA177186 itself. Looking down to lower significance levels there is nothing similar in the field of this target, within $\pm$500 kms$^{-1}$ of the systemic velocity of the target. Therefore this is an isolated, offset CO source. It is possible that the line may be from another, unidentified transition, in which case the offset source distance is unknown and it may be just a background or foreground source along a similar line-of-sight to GAMA177186. However, the observed emission line is unlikely to be anything other than $^{12}$CO(2-1) because other lines that might fall into the band are generally much weaker.
There is no optical image at the offset source location in Figure 1. \\

GAMA 272990:
Continuum image shows no detection. CO data cube seen in Figure 3 shows that emission is clearly detected
(in velocity range 11683 to 12039 km/s). 
It appears consistent with a rotating molecular gas disk of about 10\arcsec\, (8.6 kpc) in diameter.\\

GAMA 622305:
Continuum image shows no detection at the centre, but two serendipitous detections further out $\sim$13\arcsec\, away, to the north and north-west. These two serendipitous detections do not show any optical counterparts in Figure 1.
In the CO data cube there is a very weak, slightly offset detection ($\sim$2.9\arcsec) to the west, between velocities 12166 and 12257 km/s. This is only $\sim$32km/s smaller than the expected velocity for GAMA622305, if the observed emission line is $^{12}$CO(2-1).  Moment maps and the spectrum are shown in Figure 4 right, illustrating the source location and moderate rotation in the offset source.
This weak CO source is too offset to be centred on GAMA622305, unless the optical position is unusually inaccurate (by 2.9\arcsec) or the molecular gas in GAMA622305 is offset from the optical centre of the galaxy. Looking down to lower significance levels there is insufficient data to tell if there is any additional low level CO emission in the field, within $\pm$50 kms$^{-1}$ of the systemic velocity of the target. Therefore, for this offset CO source we cannot rule out the possibility that there may be additional such clumps associated with the target galaxy, below our current detection limit. There is no optical source or ALMA continuum emission centred on this offset CO emission. The CO rotation curve in Figure 4 (lower right) suggests a slowly rotating gas disc seen approximately edge on. If it is not associated with GAMA622305 then we cannot tell if it is a foreground or background source without an accurate emission line identification. If it is $^{12}$CO(2-1) that we are detecting then this offset source may be a molecular gas-rich dwarf companion galaxy to GAMA622305, which remains undetected in optical light.\\

GAMA 622429:
Continuum image shows an unresolved central detection. CO data cube seen in Figure 3 shows that emission is clearly detected (in velocity range 11434 to 12146 km/s) 
and is much more extended than the continuum emission 
It is a rotating molecular gas disk of about 11\arcsec\, (9.3 kpc) in diameter.  There is also evidence for some kinematical disturbance, from the twist seen in the velocity field in GAMA622429, see plots on the right in Figure 3. Also apparent in the spectrum is a central peak, possibly associated with the centrally concentrated emission source apparent in the continuum image (Figure 5). As discussed in Section 5.2, GAMA622429 has a Seyfert 2 AGN, which may be contributing to the central emission.\\

 In summary, three of the five galaxies have clearly associated CO emission, 
with the molecular emission aligned along the galaxy's apparent optical major axis in each case (see Table 2). Channel maps for these three galaxies are given in Appendix A. 
 Only two of the galaxies have continuum emission detected centred on the target galaxies,
however this is not as extended as the CO emission. 

Total fluxes, extents and position angles for all detections are detailed in Table 2. 
For measuring fluxes we selected apertures by incrementing the aperture until no additional flux was apparently detected. This allows for inclusion of low level, integrated flux that is not clearly visible by eye. These
$^{12}$CO(2-1) fluxes (CO$_{Flux}$) are from the aperture regions indicated (CO$_{Ap}$, ellipse major and minor axes).
CO extents (CO$_{Extent}$) are estimated by eye from the data visible in the clean data cubes, derived from CASA analysis. The extent of detectable CO emission can cover up to twice the LAS, for ordered rotation. Position angles of the extended CO emission (CO$_{PA}$) are given in degrees from north through east. Continuum emission (Cont$_{FluxDensity}$) is given either as a detection or as a 3$\sigma$ upper limit within the given aperture region (Cont$_{Ap}$, ellipse major and minor axes). Aperture regions for the continum detections are $\sim$twice the SAS, for point sources and slightly larger for the continuum source detected near GAMA177186, whilst for continuum upper limits we have approximated the area of the CO detections shown in Figure 3, assuming that dust may follow molecular gas distributions, or optical extent (for GAMA622305). Continuum upper limits are estimated, from the continuum noise per beam (Noise$_{rms}$) and the number of beams (Nbeams) in the elliptical Cont$_{AP}$ area, as a 3 sigma upper limit of (3 $\times$ sqrt(Nbeams) $\times$ Noise$_{rms}$). Errors account for random errors and 6 percent calibration source errors added in quadrature (see: https://almascience.eso.org/sc/). Serendipitous source fluxes are also given where they occur within the ALMA field, specified by their offset from the field centre in arcseconds. Continuum noise levels are given (Noise$_{rms}$). Optical r band major axes position angles (Optical$_{PA}$) for the five target galaxies are given (from NED).

\begin{table*}
\setlength{\tabcolsep}{5pt}
	\centering
	\caption{Derived properties of observed ETGs, including CO fluxes, extents and position angles, continuum flux densities (detections or limits), continuum noise levels and optical position angles. Serendipitous detections are also listed. For further details see Section 4.
}
	\label{tab:table2}
	\begin{tabular}{rcccccccc} 
		\hline
ETG &       CO$_{Flux}$ & CO$_{Ap}$  & CO$_{Extent}$  & CO$_{PA}$ &  Cont$_{FluxDensity}$ & Cont$_{Ap}$ & Noise$_{rms}$   & Optical$_{PA}$ \\
             &    (Jy km/s)  & (\arcsec)   & (diam \arcsec) &  (deg.)  &  (mJy)            &   (\arcsec)      & (mJy/beam) & (deg.)
\\		\hline
{\bf GAMA64646}           & 33.3 $\pm$2.5  & 36 x 12    & $\sim$18  &  $\sim$65 &  $<$0.977 $3\sigma$  &  15 x 7      & 0.025   &  60 \\
{\bf GAMA177186}          & -              &            &           &           &  0.516$\pm$0.077     &  1.3 x 1.3   & 0.043   &  23 \\
(3.72\arcsec,2.40\arcsec) & 2.2$\pm$0.3    & 2.4 x 2.4  &           &           &  2.080$\pm$0.157     &  1.8 x 1.8   & & \\
{\bf GAMA272990}          & 20.4$\pm$1.3   & 17 x 8.4   & $\sim$12  &   $\sim$3 &  $<$0.586 $3\sigma$  &  10 x 7      & 0.020   &    3 \\
{\bf GAMA622305}          & -              &            &           &           &  $<$1.069 $3\sigma$  &  10 x 7      & 0.034   &  145 \\
(2.80\arcsec,-0.6\arcsec) &  0.52$\pm$0.10 & 3.6 x 1.8  &           &   $\sim155$        &   -                  &              & & \\
(1.27\arcsec,12.61\arcsec)&  -             &            &           &           &  1.070$\pm$0.126     &  1.3 x 1.3   & & \\
(12.36\arcsec,4.33\arcsec)&  -             &            &           &           &  0.637$\pm$0.116     &  1.3 x 1.3   & & \\
{\bf GAMA622429}          & 59.9$\pm$3.8   & 18 x 9     & $\sim$11  & $\sim$138 &  1.100$\pm$0.085     &  1.3 x 1.3   & 0.033   &  137 \\
\hline
	\end{tabular}
\end{table*}

The serendipitous ALMA sources detected in the fields for GAMA177186 and GAMA622305 show no evidence of optical counterparts in Figure 1, neither in the SDSS nor deeper KiDS images.

\section{Discussion}
ALMA observations reveal a lack of apparent links between the molecular gas and the dust seen in emission at 1.3mm, in these five sources. Explanations for these results are discussed in this section on a case by case basis.

\subsection{CO}
Size and flux estimates are given in Table 2 and moment maps are plotted in Figure 3, where detected. The CO emission is extended where detected in GAMA64646, 272990 and 622429. There is clear rotation in these cases. In the case of GAMA64646 the extent of CO emission exceeds the LAS per channel for our ALMA observations. The CO disks detected here are generally larger than were initially expected, given typical molecular gas sizes observed in some nearby ETGs (e.g. Davis et al. 2013) and the lack of spatial extent to the H-ATLAS dust detections in our target galaxies. Alignment of the CO kinematic and photometric axes with the optical axes suggests relaxed CO disks. However, we do see some evidence for slight kinematic disturbance in GAMA622429, at a similar level to that seen in ETGs with known minor mergers (van de Voort et al. 2018). In Table 2 we use the spatial extent of CO to guide possible upper limits on detectable continuum emission. 

Molecular gas masses were estimated from the integrated line flux in the moment zero maps.
We use equation 3 from Solomon \& vanden Bout (2005) to estimate the luminosity (L21) of the CO(2-1) line in units of (K km s$^{-1}$ pc$^2$), from the measured integrated CO(2-1) line flux at a rest frequency of $\nu_{rest}=230.538$ GHz. The integrated luminosity in any CO line, in units of (K km s$^{-1}$pc$^2$) is (Solomon \& vanden Bout 2005):  

\begin{equation}
 L_{CO}=3.25 \times 10^7(S_{CO}\Delta v) (\nu_{rest})^{-2} D_L^2 (1+z)^{-1}
\end{equation}

Where ($S_{CO}\Delta v$) is the integrated line flux in (Jy km/s) (given in column 2 of Table 2 for the current targets); $D_L$ is the luminosity distance in Mpc and $z$ is the redshift. Here we have taken into account that $\nu_{rest}/\nu_{obs}=(1+z)$.

The CO intensity ratio between transitions is assumed to be r$_{21}$=CO(2-1)/CO(1-0)$\sim 1.0 \pm 0.5$ (from Young et al. 2011), which is the same as L21/L10, where L21 and L10 are the CO(2-1) and CO(1-0) luminosities in units of (K km s$^{-1}$ pc$^2$). We assume a CO-to-H$_2$ conversion factor X$_{CO}=3\times 10^{20} \pm 1\times 10^{20}$ cm$^{-2}$ (K km s$^{-1}$)$^{-1}$ (from Young et al. 2011). This corresponds to a mass-to-light ratio of $\alpha_{co}=4.8$ ($M_{\odot}$/(K km s$^{-1}$pc$^2$)), times a factor 1.36 to allow for helium mass. Hence the molecular gas mass can be estimated from:

\begin{equation}
 M_{mol}=1.36 \alpha_{co} {\rm{L21}}(r_{21})^{-1}
\end{equation}

We thus estimate the total molecular gas masses ($M_{mol}$) and uncertainties from the measure integrated CO line fluxes given in Table 2. Using dust masses and uncertainties estimated from the latest H-ATLAS data, based on multiwavebands fits using MAGPHYS (from GAMA Data Release 3, MagPhys v06 catalogue{\footnote {http://www.gama-survey.org/dr3/data/cat/MagPhys/v06/ }}), we can then estimate molecular gas-to-dust mass ratios. These results are given in Table 3. Dust masses and their uncertainties given in Table 3 supercede previous estimates (such as those from Agius et al. 2013) because they include the most recent measurements of errors and improved calibration of the H-ATLAS data (Valiante et al. 2016). Errors shown for M$_{mol}$ are from CO flux errors only. A further 33 percent uncertainty in M$_{mol}$ comes from uncertainty in X$_{CO}$ and 50 percent uncertainty from CO(2-1)/CO(1-0) ratio. These latter two uncertainties are scaling factors, rather than absolute uncertainties on M$_{mol}$ molecular gas masses. 
These calculations assume that the molecular gas is optically thin across the galaxy.

\begin{table*}
\setlength{\tabcolsep}{4pt}
	\centering
	\caption{Dust and molecular gas mass estimates for the observed ETGs. Other variabes are also listed, from GAMA DR3 MAGPHYS fits. Colons are used for highly uncertain values, as indicated in the Notes column.  See Section 5.1 for details.}
	\label{tab:table1}
	\begin{tabular}{lccccccccccc} 
		\hline
ETG    & MAGPHYS     & MAGPHYS    &         & MAGPHYS &              &         &                       &                       &         & (M$_{mol}$/SFR)     & \\
GAMA   & M$_*$       & M$_d$      &  $\pm$  & SFR     &  M$_{mol}$    & $\pm$   & ${{\rm M}_{mol}}\over{{\rm M}_*}$  & ${{\rm M}_{mol}}\over{{\rm M}_d}$  & $\pm$  & t$_{depl}$           & Notes\\
ID     & (M$_\odot$) & (M$_\odot$) &         & (M$_\odot$/yr) &  (M$_\odot$)  &         &                 &                       &        & (yr)                & \\  
		\hline
 64646 & 7.80E+10   & 3.3E+7     & 4.2E+6   & 1.066 & 3.2E+9   & 2.4E+8     & 0.0410            &  98  &  14 & 3.0E+9  &\\
177186 & 1.68E+10   & 1.9E+7:    & 4.0E+6   &  -    & 3.5E+8   & 4.8E+7     &   -               &  18: &   5 &  -      & Offset contamination\\
272990 & 3.34E+10   & 2.3E+7     & 3.3E+6   & 0.608 & 2.4E+9   & 1.6E+8     & 0.0719            & 106  &  17 & 3.9E+9  &\\
622305 & 4.27E+10   & 3.9E+7     & 6.8E+6   & 0.490 & 6.6E+7:  & 1.3E+7     & 0.0015:           &   2: &   1 & 1.3E+8: & Offset source CO\\
622429 & 2.97E+10   & 4.2E+7     & 3.4E+6   & 4.245 & 7.1E+9   & 4.4E+8     & 0.2391            & 168  &  17 & 1.7E+9  &\\
		\hline
	\end{tabular}

\end{table*}

For our three ETGs with clearly detected associated CO gas (GAMA64646, GAMA272990 and GAMA622429) the molecular gas masses are large, with a few$\times 10^{9}$M$_\odot$ of molecular gas and a molecular gas mass that is a quarter of the stellar mass in the case of GAMA622429, for example. This is at the top end of the range seen in local spiral galaxies of comparable stellar mass, which typically have between a tenth and a hundredth of their mass in molecular gas (e.g. Papovich et al. 2016, their figure 3; Saintonge et al. 2017a). From the mass selected sample of nearby galaxies in Saintonge et al. (2017a, their figure 10), galaxies with similar masses to our three ETGs have a mean molecular gas fraction of about a twentieth (from stacking) of their stellar mass, but with more than an order of magnitude in range. Optically selected ETGs from the ATLAS$^{3D}$ survey (Young et al. 2011) have a wide range of molecular gas masses (Log(M$_{mol}$) ranging from 7.1 to 9.29 in solar units), but with little dependence on stellar mass (Young et al. 2011, their figure 8). Therefore, at the typical stellar mass of our three ETGs (Log(M$_* \sim 10.6$M$_\odot$), the ATLAS$^{3D}$ ETGs have less than a twentieth of their mass in molecular gas. Our three ETGs are at the high end or exceed the molecular gas mass fractions observed in ATLAS$^{3D}$ ETGs.
The Milky Way Galaxy has only a total molecular gas mass of 6.5$\times 10^8$M$_\odot$ (Roman-Duval et al. 2016). Therefore these three galaxies (GAMA64646, GAMA272990 and GAMA622429) have large reservoirs of molecular gas, for their stellar mass, but their optical morphology is that of early-type galaxies.

For these three ETGs the molecular gas-to-dust mass ratios are on average M$_{mol}$/M$_d$=124 (from Table 3). This is comparable to the average ratio of $\sim$96 (ranging from 23 to 467) found for seventeen dust lane ETGs in Davis et al. (2015, their table 3), with molecular gas masses based on IRAM observations. 
We compared to other galaxy types as well. Star forming dwarf galaxies, with low metallicity are the main galaxy types known to have larger gas-to-dust mass ratios (R\'{e}my-Ruyer et al. 2014), but not as much absolute molecular gas mass as seen here. Remy-Ruyer et al. published gas masses for a range of galaxy types (mostly late-type galaxies) and metallicities. For 20 (non-dwarf) galaxies with 12+log(O/H)>8.5, their mean molecular mass is M$_{mol}$=$9.1\times 10^9$M$_\odot$ and their mean molecular gas-to-dust mass ratio is M$_{mol}$/M$_d$=98 (ranging from 9 to 302). We assumed their metal dependent CO-to-H$_2$ conversions and a factor 1.36 to include helium in estimated molecular gas masses from Remy-Ruyer et al. (2014). The Herschel reference Survey, including a range of galaxy types, shows a similar large range in molecular gas-to-dust mass (Cortese et al. 2016). The gas-to-dust mass ratios in our three dusty ETGs are therefore similar to those in late-type galaxies, although all samples show a large range.  
 
Four of the five targets have UV detections from GALEX satellite. Their star formation rates (SFRs) estimated in GAMA DR3 are in the range 0.49 to 4.24 M$_\odot$/year, with large uncertainty depending on the method used (e.g. MAGPHYS fits, H$\alpha$, UV and/or IR fluxes). The typical level is comparable to the Milky Way (e.g. 1.25 M$_\odot$/year given in Popescu et al. 2017) and much less than in starburst galaxies. GAMA177186 has weaker evidence for UV emission and a much lower SFR. With these rates of star formation and stellar masses, our ETGs fall below or within the band defining the galaxy main sequence from Saintonge et al. (2017a, their figure 8), except for GAMA622429, which falls above it. Thus for their stellar masses and SFRs they have large molecular gas mass fractions compared to the sample of Saintonge et al. (2017a). In fact, our three galaxies with large reservoirs of molecular gas lie close to the post-starburst galaxies shown in the recent study by French et al. (2018, their figure 2). Depletion times are estimated as $t_{depl}=M_{mol}/SFR$ and are listed in Table 3, with typically a few $\times 10^9$ years. These relations will be studied in more detail in a future paper.

\begin{figure*}
	\hspace{-0.5cm}\includegraphics[width=182mm]{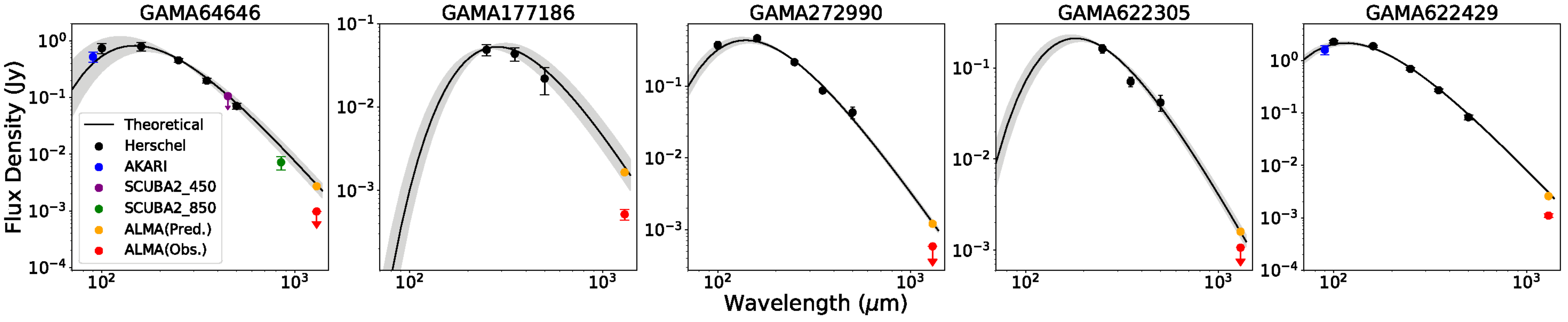}
    \caption{SEDs for our five target galaxies, showing H-ATLAS fluxes from data release 1 (Valiante et al. 2016), AKARI 90$\mu$m and SCUBA-2 850$\mu$m fluxes where available (see Section 5.2), together with the H-ATLAS temperature fits (including uncertainties in grey) from Agius et al. (2013), assuming $\beta=2$ and normalised here to the latest 250$\mu$m flux estimates (from Valiante et al. 2016). ALMA predictions for Cycle 3 are shown in orange and ALMA continuum flux observations or 3$\sigma$ upper limits as given in Table 2 are shown in red.}
    \label{fig:SEDs_figure}
\end{figure*}

\begin{figure*}
    \hspace{-0.8cm}
    \includegraphics[width=184mm]{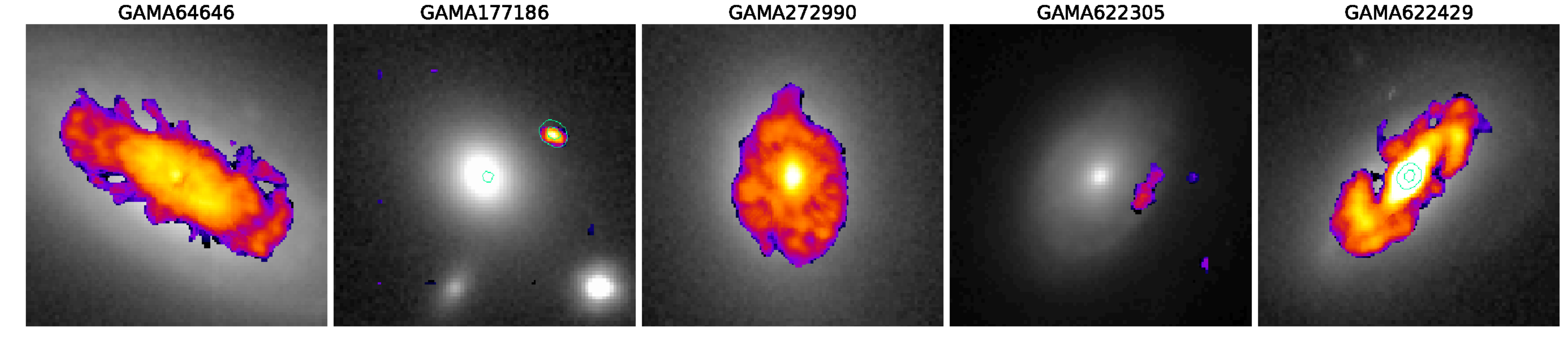}
    \caption{Composite images for the central 17\arcsec\, by 17\arcsec\, regions of our five target galaxies, centred on the optical position of the galaxies. Greyscale plots show KiDS r-band images with logarithmic scaling. Coloured overlays show the ALMA moment zero line emission as in Figures 3 and 4 above, with colour representing intensity in this figure. Green contours indicate the ALMA continuum emission near the centre of GAMA177186, its offset source at 4.4\arcsec\' north west of the target, and near the centre of GAMA622429.}
    \label{fig:Overlays_figure}
\end{figure*}

Considering possible origins for the molecular gas, the kinematics of two of the galaxies looks undisturbed (GAMA64646, GAMA272990), indicating that either the gas is formed in-situ from stellar mass loss, or the gas was accreted more than a few hundred Myrs ago. If accreted, then the large amounts (M$_{mol} \sim$ few $\times 10^9$M$_\odot$) indicate a massive accretion event and/or an extended duration of accretion to build up those molecular gas reservoirs. In the case of GAMA177186, GAMA622305 and GAMA622429 it might be that the gas has recently been accreted externally, because the CO is kinematically and morphologicallly disturbed.

\subsection{Continuum}
Comparisons of ALMA continuum source detections and limits versus expected extrapolations from H-ATLAS fluxes, and origins of ALMA fluxes, are discussed in this section. 

 Spectral energy distributions (SEDs) are plotted in the submm range in Figure 6. These SEDs include data from H-ATLAS (up to 5 wavebands) in each ALMA target, AKARI (for two galaxies), SCUBA-2 (for one galaxy) and measured ALMA continuum fluxes or three sigma upper limits for detectable flux from these ALMA data. 
We checked for emission from AKARI (Yamamura et al. 2010), which was in operation a few years before Herschel and found emission at 90$\mu$m for GAMA64646 and GAMA622429, which was not significantly offset from the galaxy. 
AKARI catalogue flux densities were obtained via IRSA{\footnote{NASA/IPAC Infrared Science Archive iras.ipac.caltech.edu}} and were colour corrected 
based on the temperature and emissivity index used to plot the SEDs in Figure 6. These AKARI flux densities are plotted in Figure 6 and agree quite well with H-ATLAS data and SED fits from H-ATLAS. The SCUBA-2 data point for GAMA64646 are from Saintonge et al. (2017b).

The line and grey regions plotted in Figure 6 correspond to the best fitting single modified black body temperature from Agius et al. (2013), normalised here to the most recent H-ATLAS 250$\mu$m flux density (from data release 1, Valiante et al. 2016), indicating the extrapolated prediction to ALMA at 1.3mm. A conservative approach was taken by fixing $\beta=2$ in the modified black body to avoid overestimating fluxes at 1.3mm.  In general, we detect much less continuum flux than expected, given the strong H-ATLAS detections and ALMA CO detections of molecular gas.  

 Figure 6 shows that the extrapolated predictions from H-ATLAS data are generally higher than the observed (or upper limits to) ALMA fluxes. GAMA64646 was also detected with SCUBA-2 as part of the JINGLE survey (Saintonge et al. 2017b) and their 850$\mu$m flux agrees quite well with the H-ATLAS fitted SED. If the dust known to be in GAMA64646 is distributed as the molecular gas detected in this work, then the continuum emission will have been largely resolved out in our ALMA observations. This is probably the reason why we do not detect the dust in GAMA64646 with these ALMA observations. However, the Herschel PACS detections for GAMA64646 show a compact rather than extended dust distribution. Cold dust emission might be more extended than warmer dust detected in PACS because the warm dust is more sensitive to heating sources (e.g. see Popescu \& Tuffs 2013, who show that warm dust is more confined than cold dust in our own Galaxy). For GAMA64646 we also attempted to taper the contributions from larger baseline to reduce sensitivity to small scale structures, however, this also led to no continuum detection in the cleaned image. 
Continuum flux will be resolved out of the ALMA observations on scales larger than the LAS. Estimates of upper limits can only constrain that proportion of continuum flux that could be detected by the particular ALMA array being used. Other authors have sometimes detected faint, extended 1mm dust emission in more nearby ETGs (e.g. Boizelle et al. 2017; Davis et al. 2017). This lends support to the idea that continuum is resolved out in some of these current ALMA observations.

For the ALMA observations of our other two CO detected sources (GAMA272990 and GAMA622429), their CO extents are comparable to the LAS. Therefore it is possible that some continuum emission has been resolved out if dust is distributed as the gas, but it is surprising that no dust is detected at all in the ALMA observations of GAMA272990. If dust is distributed as the detected CO in these two cases, this is more extended than expected and the low surface brightness would help to explain the lack of extended sources in the ALMA continuum images (Figure 5). Future interferometric observations of dust in ETGs need to take into account these larger molecular gas extents, meaning that longer exposures are needed to map the dust distributions in ETGs, even for those known to contain large dust masses (see also van de Voort et al. 2018, their table 3). Exact expectations will depend on how the dust emission is distributed. Assuming a sersic index of 1 for the surface brightness profile, and continuum flux at 1.3$\mu$m of $\sim 2.5 mJy$ (see predictions in Figure 6, for GAMA64646, for example), then for dust extended by 5\arcsec\, by 5\arcsec\, the central beam detections would be 3.3$\sigma$ above the measured noise level of 0.025 mJy/beam. If the dust in GAMA64646 is extended as the CO (15\arcsec\, by 7\arcsec), then the observation is surface brightness limited and the central flux would not reach 1$\sigma$. This explains the weak or non-detections of continuum flux in our 3 target galaxies with strong CO disks detected, if the dust is distributed like the molecular gas. An alternative possibility is collisional heating of the dust, if these galaxies possess a hot ISM. In this case the dust may not be distributed like the molecular gas. Investigation of this possibility would need X-ray observations and detailed modelling. 

Future detailed simulations will aid in understanding these results in more detail. There might be a turndown, corresponding to a change in emissivity, towards low frequencies in some of these ETG spectra, however observations at intermediate wavebands (e.g. SCUBA-2) are needed to determine if this is the case.
For GAMA622429 a point source is detected, which may be associated with dust around the centre of this galaxy, where there is compact radio emission and star formation (Best \& Heckman 2012). This galaxy is classified as a Seyfert 2 in NED, which had been missed in our previous emission line analysis of the GAMA data (Agius et al. 2013). It is listed as a radio-quiet (i.e. no strong radio jets) in G{\"u}rkan et al. (2015). GAMA622429 shows signs of a dust lane in the KiDS image (Figure 1) and the detected continuum emission may be part of this more extended, but weaker dust distribution. 

Composite plots of the central 17\arcsec by 17\arcsec are given in Figure 7, which shows the relative locations of the KiDS r-band optical emission (background greyscale), ALMA line emission (false colour overlay) and  ALMA continuum emission (green contours at 0.05 and 0.5 of the peak flux), detected in our five target ETGs.

For GAMA177186 there is a brighter continuum source only 4.4\arcsec\, away, which could have contaminated the H-ATLAS flux measurements. In the ALMA continuum data the offset source is four times as bright as the point source centred on GAMA177186 (see Figure 5 and measured fluxes in Table 2). If we include the offset source flux with the GAMA177186 flux at 1.3mm, then the total continuum flux is consistent (within errors) with our predicted ALMA value (shown as an orange point in the second panel of Figure 6). Therefore we estimate that this offset source may have contributed $\sim$80 percent of the H-ATLAS flux, depending on the temperature of the offset source. Single modified backbody parameters are not really representative of the dust properties in the case of GAMA177186 shown in Figure 6 because the spectral energy distribution is a blend of emission from two different sources. It was not possible to resolve these two sources with Herschel, but ALMA data clearly resolves them.

For GAMA622305 there is no ALMA CO or continuum emission centred on the target, but a weak offset CO detection is only 2.9\arcsec\, away. Might undeteced dust, associated with this offset source, have contaminated the H-ATLAS data for GAMA622305? Very little contamination would be expected from such a weak CO source, unless the dust-to-gas ratio is unexpectedly high.
Two other serendipitous continuum detections at $\sim$13\arcsec\, away might contribute some contaminating flux to the H-ATLAS detection of GAMA622305, however, if they dominated the H-ATLAS fluxes then an offset of $\sim$13\arcsec\, would be expected between the optical and H-ATLAS centres, which is not seen (see Figure 1). It is possible that sources with different temperatures might dominate in different submm wavebands, confusing the H-ATLAS bands; however this seems unlikely in the case of GAMA622305 because of the lack of detection at H-ATLAS 250$\mu$m of the two seredipitous ALMA sources. Therefore we conclude that these two other serendipitous continuum detections in this field are too far away to have contaminated the H-ATLAS 250$\mu$m fluxes.

\section{Conclusions}

ALMA band 6 observations were obtained for five early-type galaxies that had large dust mass estimates from H-ATLAS detections. These reveal the spatial and kinematic distributions of their molecular gas. We detect rotating molecular gas discs in three out of five of the target ETGs, plus weak offset CO emission in the other two target fields. For our three CO disk detected ETGs (GAMA64646, GAMA272990 and GAMA622429), their typical M$_{mol}$/M$_d$ is $\sim$ hundred (124 on average), which is similar to that seen in most ETGs with optically detected dust lanes. These three may have a cold ISM dominated by their molecular gas mass. We find $\sim$ few $\times 10^{9}$M$_\odot$ of molecular gas, which is unusually large for early-type galaxies in general and larger than typically expected in late-type galaxies of the same stellar mass. 

Only in two of these five ETGs (GAMA177186 and GAMA622429) do we detect continuum emission centred on the galaxy and one of those has an additional emission sources at only 4.4\arcsec\, away. This is clearly separated in the ALMA observations, but was not resolved in the H-ATLAS observations. Therefore one of our five targets (GAMA177186) may be affected by contamination in the Herschel data. For the other four targets there is no evidence from these ALMA observations for contamination (in the H-ATLAS detections) by a continuum submm source other than the target galaxies themselves. Additional sources serendipitously detected in the ALMA data are too far away ($\sim$13\arcsec) or too weak to be candidates for contamination in the Herschel 250$\mu$m data.

Spectral energy distribution plots reveal a lack of ALMA dust detections in these data when compared with predicted extrapolations from observations at shorter wavelengths, including AKARI, H-ATLAS and SCUBA-2 data. The low level of dust detections is also surprising given the clearly detected CO emission in three cases. In one case (GAMA64646) the continuum emission has been resolved out because the full extent of detected CO is more than 18\arcsec (see Figure A1). Surface brightness limits also affect these ALMA detections. 

The largest angular size detectable from our ALMA observations is typically 11\arcsec, which is comparable to the largest diameters observed across the detected CO emission from the molecular discs observed in two of our targets (GAMA272990 and GAMA622429). Therefore, if the dust distribution is as extended as the CO emission we may have resolved out some of the continuum flux in our ALMA maps and the rest must be below the surface brightness limit. However, it is still surprising that we see so little continuum flux from our ALMA observations because their dust distributions are not very extended in the H-ATLAS PACS detections at shorter wavelengths. The small extent of dust seen in PACS data may be compatible with the lack of dust detected in the ALMA observations if cooler dust is more extended than warmer dust, as seen in some galaxies. Detailed SED and CO emission modelling will be carried out for these targets in a future paper, taking account of what we have learnt from these ALMA observations.

We have shown that while molecular gas is quite easily mapped in these ETGs, it will take much deeper observations to map the distribution of their dust emission. Given the heterogeneous findings in this small sample, larger samples need to be spatially resolved to understand the dust-to-gas mass ratios and distributions more accurately in ETGs.

\section*{Acknowledgements}

This paper makes use of the following ALMA data: ADS/JAO.ALMA$\#$2015.1.00477.S. ALMA is a partnership of ESO (representing its member states), NSF (USA) and NINS (Japan), together with NRC (Canada), MOST and ASIAA (Taiwan), and KASI (Republic of Korea), in cooperation with the Republic of Chile. The Joint ALMA Observatory is operated by ESO, AUI/NRAO and NAOJ.
The KiDS images are based on data products from observations made with ESO Telescopes at the La Silla Paranal Observatory under programme IDs 177.A-3016, 177.A-3017 and 177.A-3018, and on data products produced by Target/OmegaCEN, INAF-OACN, INAF-OAPD and the KiDS production team, on behalf of the KiDS consortium. OmegaCEN and the KiDS production team acknowledge support by NOVA and NWO-M grants. Members of INAF-OAPD and INAF-OACN also acknowledge the support from the Department of Physics \& Astronomy of the University of Padova, and of the Department of Physics of Univ. Federico II (Naples).
The composite KiDS images shown in this paper were formed by Lee Kelvin at the ARI in Liverpool John Moores University.  
JINGLE is a large program led by A. Saintonge, which is undertaking submm obsevations of galacies with the JCMT.
Thanks to Megan Argo and to an anonymous referee for their constructive comments.









\appendix

\section{Channel maps for CO emission}

Channel maps for GAMA64646 from the data cube covering CO emission reveal molecular gas extending out to the LAS per channel (see Figure A1). Given the systematic velocity shift with position this means that we can see emission from a range of at least 18'' across. This size is larger than anticipated. If the dust (detected in H-ATLAS and SCUBA-2) follows the molecular gas in this galaxy, and is diffusely distributed, then our current ALMA observations will have resolved out the dust. Observations from a more compact array configuration are needed to find out whether there is dust on larger scales or not. 

Channel maps for GAMA272990 and GAMA622429 are shown in Figures A2 and A3, where the CO emission is seen to be confined to a full extent that is within the LAS of 11\arcsec\, across.

\begin{figure*}
	\includegraphics[width=145mm, angle=0]{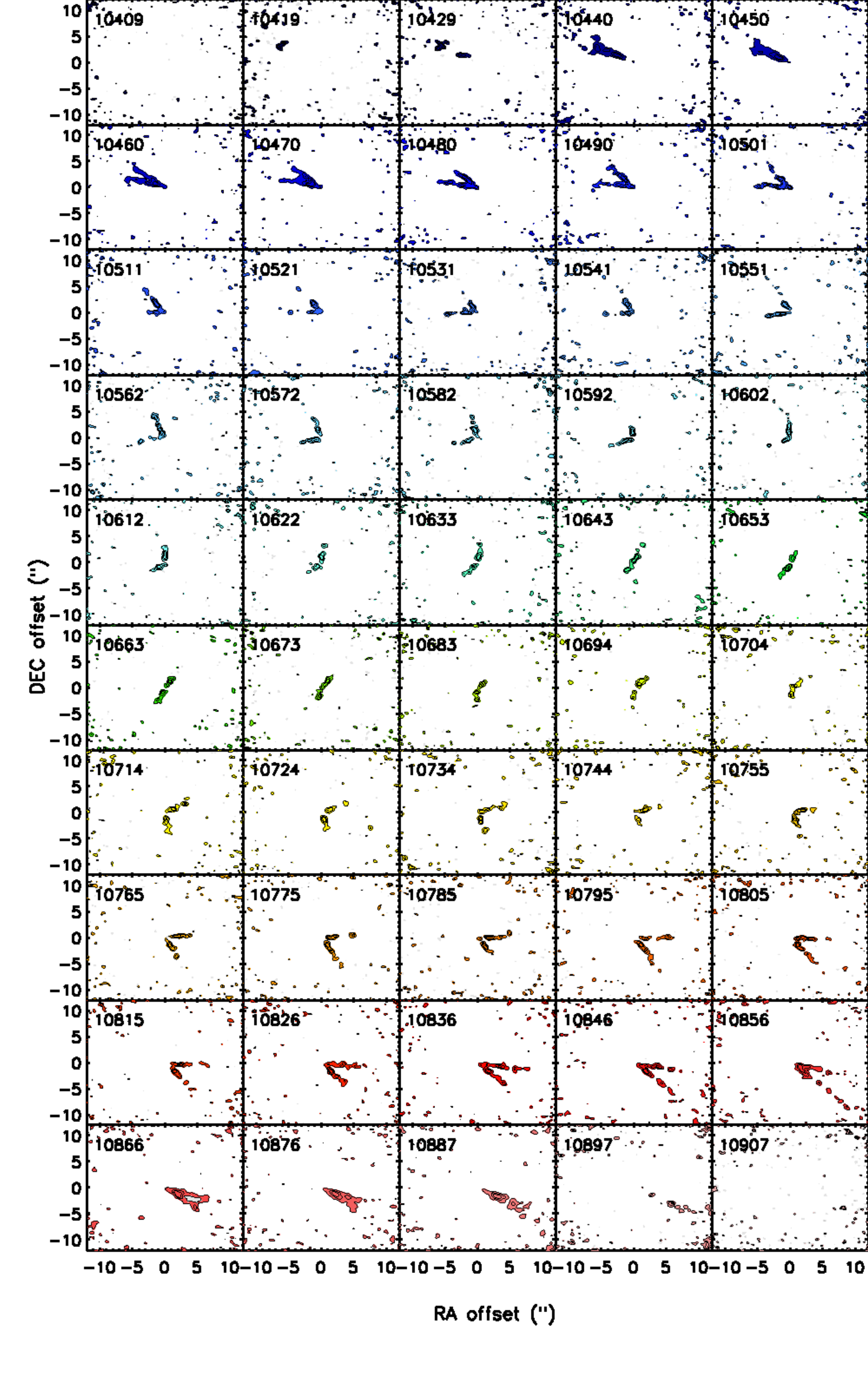}
   \caption{Channel maps for GAMA64646 showing the extended CO emission from this galaxy, across its detected velocity channels. The central velocity is shown top left in each channel map, in km s$^{-1}$.
}
    \label{fig:G64646_channelmaps}
\end{figure*}

\begin{figure*}
	\includegraphics[width=160mm, angle=0]{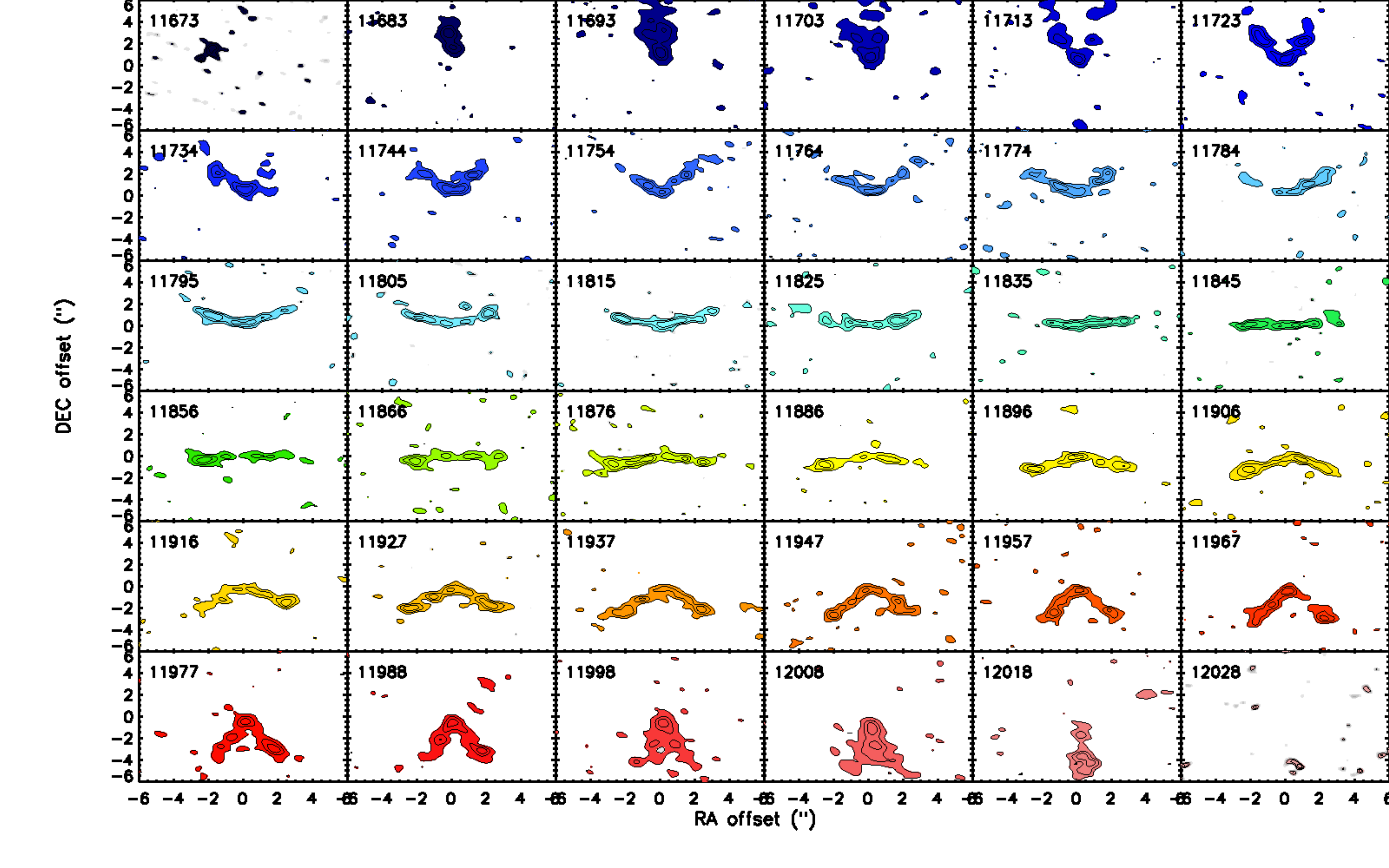}
   \caption{Channel maps for GAMA272990 showing the extended CO emission from this galaxy, across its detected velocity channels. The central velocity is shown top left in each channel map, in km s$^{-1}$.
}
    \label{fig:G272990_channelmaps}
\end{figure*}

\begin{figure*}
	\includegraphics[width=160mm, angle=0]{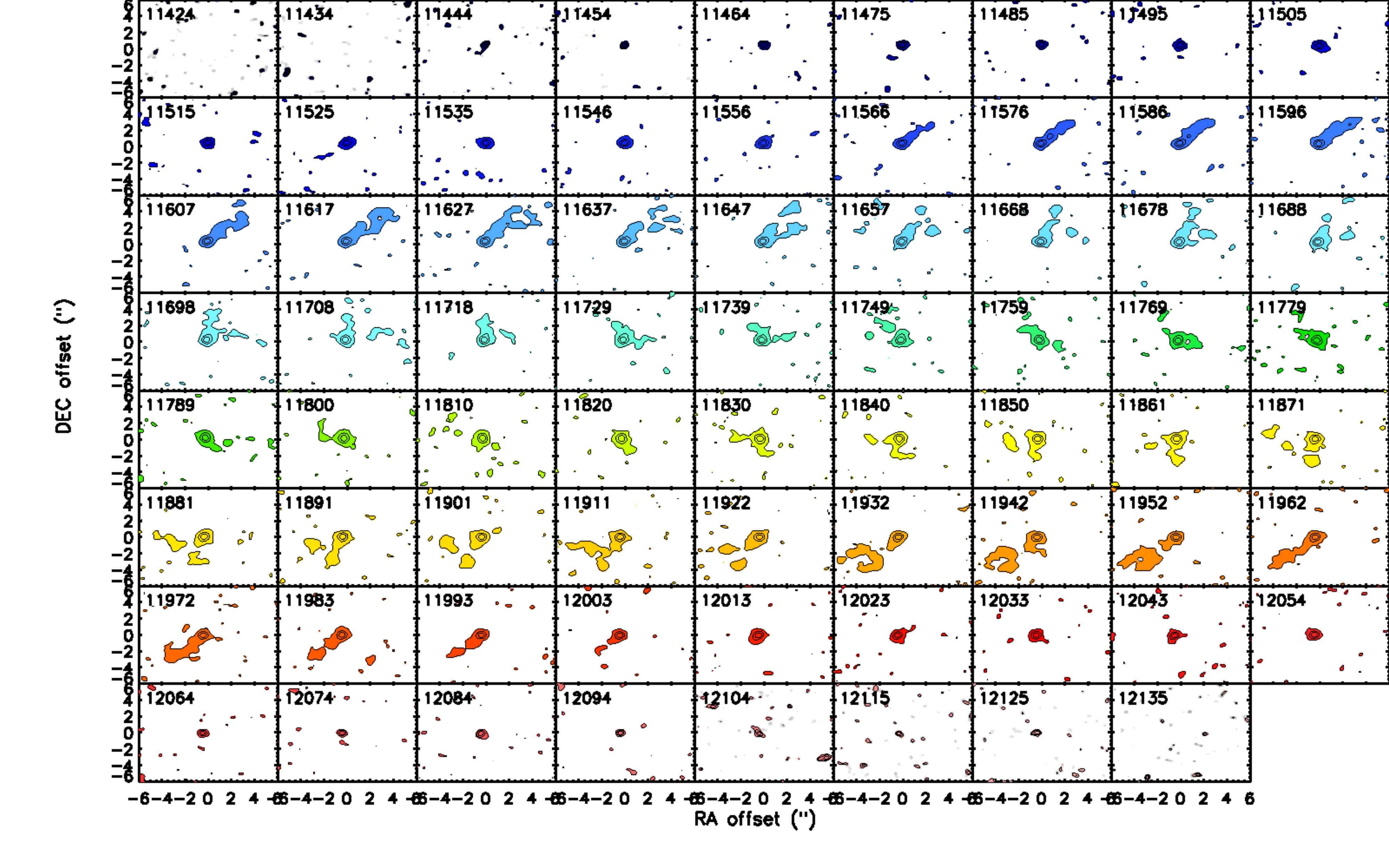}
   \caption{Channel maps for GAMA622429 showing the extended CO emission from this galaxy, across its detected velocity channels. The central velocity is shown top left in each channel map, in km s$^{-1}$.
}
    \label{fig:G622429_channelmaps}
\end{figure*}


\bsp	
\label{lastpage}
\end{document}